\documentstyle[aps,prl,twocolumn,epsfig]{revtex}

\begin{document}

\twocolumn[\hsize\textwidth\columnwidth\hsize
\csname@twocolumnfalse%
\endcsname

\draft

\title{Ground-state properties of a dilute Bose-Fermi mixture}

\author{L. Viverit$^{1}$ and S. Giorgini$^{2}$}

\address{$^{1}$Dipartimento di Fisica, Universit\`a di Milano, via Celoria
16, I-20133 Milan, Italy}
\address{$^{2}$Dipartimento di Fisica, Universit\`a di Trento, \protect\\
and Istituto Nazionale di Fisica della Materia, I-38050 Povo, Italy}

\maketitle

\begin{abstract}

We investigate the properties of a dilute boson-fermion 
mixture at zero temperature. 
The ground-state energy of the system is calculated to second 
order in the Bose-Fermi 
coupling constant. The Green's function formalism is applied to obtain results for the 
effect of Bose-Fermi interactions on the spectrum of phonon excitations, the momentum 
distribution, the condensate fraction and the superfluid density of the bosonic component.
A quantitative discussion of these effects is presented in the case of mixtures with
attractive interspecies interaction.

\end{abstract}

\vskip2pc]


\narrowtext

\section{Introduction}

Recent experimental achievements in the trapping and cooling of mixtures of bosonic and fermionic 
atoms have attracted much attention on the physics of dilute Bose-Fermi mixtures in the quantum 
degenerate regime. Stable Bose-Einstein condensates (BEC) immersed in a degenerate Fermi gas have 
been realized with $^7$Li in $^6$Li \cite{SCHRECK}, $^{23}$Na in $^6$Li \cite{KETTERLE} and
very recently with $^{87}$Rb in $^{40}$K \cite{ROATI}. In particular, the $^{87}$Rb-$^{40}$K 
mixture appears to be highly interesting because of the large and negative interspecies scattering
length which favours the stability of the mixture. Current theoretical investigations of dilute 
Bose-Fermi mixtures have mainly addressed the determination of the density profiles of the two components 
in trapped systems \cite{TBF}, the problem of stability and phase separation \cite{VIVPS,PS} and 
the BEC-induced interaction between fermions \cite{II}. The effect of boson-fermion interactions
on the phonon excitation spectrum has been investigated in Refs. \cite{YIP} and \cite{PZWM}. 
Further, perturbation theory has been applied to the calculation of the ground-state energy in the 
regime of high fermion concentration by Albus {\it et al.} \cite{AGIW}, whereas the opposite regime 
of low fermion concentration has been investigated long ago by Saam \cite{SAAM}.

In the present paper we consider a homogeneous dilute mixture of a normal Fermi gas and a Bose-Einstein 
condensed gas at zero temperature.   
We calculate the ground-state energy of the system to second order in the Bose-Fermi coupling constant, 
interpolating between the high and low fermion concentration regimes, and obtain results for the dispersion 
of phonon excitations, including corrections to the sound velocity and Landau damping.
In addition to these results, we focus on the effect of boson-fermion interactions on the momentum distribution 
of the bosonic component, on the depletion of the condensate state and on the superfluid density. 
We find that the boson momentum distribution can be strongly affected by the fermions, resulting in a 
suppression of the occupation of low momentum states and in a long tail at large momenta. 
We also point out the possible occurrence of striking features caused by the Bose-Fermi coupling, such as the 
increase of the condensate fraction compared to the pure bosonic case and the superfluid density 
becoming smaller than the condensate fraction. 

The structure of the paper is as follows. In Sec. II we introduce the basic Hamiltonian of the system
within the Bogoliubov approximation. In Sec. II A we calculate the ground-state energy by direct use of
perturbation theory. In Secs. II B-D we introduce the Green's function formalism which we apply to
calculate the elementary excitation energies, the momentum distribution, the condensate fraction and the
superfluid density of the bosonic component. In Sec. III we comment on the stability of the mixture against 
phase separation and we present numerical results obtained for specific configurations of mixtures with 
attractive Bose-Fermi coupling. Finally, in Sec. IV we draw our conclusions.      
   
\section{Theory}

We consider a stable homogeneous mixture of a Bose gas and a spin-polarized Fermi gas. 
The Hamiltonian of the system can be written as the sum
\begin{equation}
H=H_F+H_B+H_{int} 
\label{ham1}
\end{equation} 
of the pure fermionic ($H_F$) and bosonic ($H_B$) Hamiltonian, and of the interaction term  
($H_{int}$) which accounts for the coupling between the two species.
If one neglects p-wave interactions in the spin-polarized Fermi gas, $H_F$ corresponds to the 
Hamiltonian of a free gas
\begin{equation}
H_F=\sum_{\bf k} \epsilon_k^F b_{\bf k}^\dagger b_{\bf k} \;,
\label{ham2}
\end{equation}
where $b_{\bf k}$, $b_{\bf k}^\dagger$ are the annihilation and creation operators of fermions and
$\epsilon_k^F=\hbar^2k^2/2m_F$ is the energy of a free particle of mass $m_F$.
The pure bosonic component is described instead by the Bogoliubov Hamiltonian
\begin{equation}
H_B=E_B + \sum_{\bf k} \omega_k \alpha_{\bf k}^\dagger\alpha_{\bf k} \; ,
\label{bog1}
\end{equation}
written in terms of the quasi-particle annihilation and creation operators $\alpha_{\bf k}$, 
$\alpha_{\bf k}^\dagger$. These operators are related to the bosonic 
particle operators $a_{\bf k}$, $a_{\bf k}^\dagger$ 
through the well-known canonical transformation 
\begin{eqnarray}
a_{\bf k}&=&u_k \alpha_{\bf k} + v_k \alpha_{- {\bf k}}^\dagger \nonumber\\
a_{\bf k}^\dagger&=&u_k \alpha_{\bf k}^\dagger + v_k \alpha_{- {\bf k}} \;, 
\label{bog2}
\end{eqnarray}
with coefficients $u_k^2=1+v_k^2=(\epsilon_k^B+g_{BB}n_0+\omega_k)/2\omega_k$ and $u_kv_k=-g_{BB}n_0/2\omega_k$. 
The elementary excitation energies obey the usual Bogoliubov spectrum
\begin{equation}
\omega_k=\sqrt{(\epsilon_k^B)^2+2g_{BB}n_0\epsilon_k^B} \;,
\label{bog3}
\end{equation} 
where $\epsilon_k^B=\hbar^2k^2/2m_B$ is the free particle energy, $n_0$ is the condensate density and 
$g_{BB}=4\pi\hbar^2a/m_B$ is the coupling constant between bosons fixed by the mass $m_B$ and the 
$s$-wave scattering length $a$ which we assume positive. The constant term
\begin{equation}    
E_B=N_B \, \left[ 4\pi n_Ba^3 + \frac{512}{15}\sqrt{\pi}(n_Ba^3)^{3/2} \right] \; \frac{\hbar^2}{2m_Ba^2}
\label{bog4}
\end{equation} 
is the ground-state energy of a dilute Bose gas expressed in terms of the gas parameter $n_Ba^3$, 
with $n_B=N_B/V$ the total density of bosons. Result (\ref{bog4}) includes the zero-point motion of the 
elementary excitations. Finally, the Hamiltonian
\begin{equation}
H_{int}= g_{BF} \int d{\bf r}\; n_B({\bf r}) n_F({\bf r})
\label{ham3}
\end{equation}
describes the interaction between fermions and bosons through the coupling constant $g_{BF}=\pm 2\pi\hbar^2b/m_R$,
which can be either positive or negative depending on whether the interactions are repulsive or attractive.
The coupling constant $g_{BF}$ is fixed by the reduced mass $m_R=m_Bm_F/(m_B+m_F)$ and by the modulus of 
the boson-fermion s-wave scattering length $b$. 
By introducing the bosonic and fermionic density fluctuation operators $\rho_{\bf k}^B=1/\sqrt{V}\int d{\bf r}\; 
e^{i{\bf k}\cdot{\bf r}}[n_B({\bf r})-n_B]$ and $\rho_{\bf k}^F=1/\sqrt{V}\int d{\bf r}\; 
e^{i{\bf k}\cdot{\bf r}}[n_F({\bf r})-n_F]$, where $n_F=N_F/V$ is the mean density of fermions, the interaction 
Hamiltonian can be rewritten in the following form
\begin{equation}
H_{int}= g_{BF}N_B n_F + g_{BF}\sum_{\bf k} \rho_{\bf k}^B \rho_{-{\bf k}}^F \;.
\label{ham4}
\end{equation}
The constant term in the above equation corresponds to the mean-field interaction energy, the second term
describes the coupling between density fluctuations in the two species.
Within the Bogoliubov approximation the bosonic density fluctuation operator can be written as a linear 
combination of quasi-particle operators
\begin{equation}
\rho_{\bf k}^B=\frac{1}{\sqrt{V}} \sum_{\bf q}a_{\bf q}^\dagger a_{{\bf q}+{\bf k}}\simeq \sqrt{n_0}(u_k+v_k)
(\alpha_{\bf k}+
\alpha_{-{\bf k}}^\dagger) \; .
\label{ham5}
\end {equation}
By using the above expression for $\rho_{\bf k}^B$, the total Hamiltonian (\ref{ham1}) takes the form
\begin{eqnarray}
H&=&E_B + g_{BF}N_B n_F + \sum_{\bf k} \epsilon_k^F b_{\bf k}^\dagger b_{\bf k} + 
\sum_{\bf k} \omega_k \alpha_{\bf k}^\dagger\alpha_{\bf k}
\nonumber \\
&+& g_{BF}\sqrt{n_0} \sum_{\bf k} (u_k+v_k)(\alpha_{\bf k}+\alpha_{-{\bf k}}^\dagger) \rho_{-{\bf k}}^F \;.
\label{ham6}
\end{eqnarray}
The Hamiltonian (\ref{ham6}) is treated in perturbation theory. We write $H=H_0+H_{int}$, where $H_0=H_F+H_B$ is the 
unperturbed Hamiltonian which is diagonal in the fermionic particle operators and in the bosonic elementary excitation 
operators, and
\begin{eqnarray}
H_{int}&=&g_{BF}N_B n_F 
\nonumber\\
&+& g_{BF}\sqrt{n_0} \sum_{\bf k} (u_k+v_k)(\alpha_{\bf k}+\alpha_{-{\bf k}}^\dagger) 
\rho_{-{\bf k}}^F
\label{ham7}
\end{eqnarray}
is the perturbation term.
Since we make use of the pseudopotential approximation for both the boson-boson and the boson-fermion
interatomic potential, the present approach is valid if both gas parameters are small, i.e. if $n_Ba^3\ll 1$ 
and $n_Fb^3\ll 1$.

\subsection{Ground-state energy}

The unperturbed ground-state energy can be readily calculated from the Hamiltonian $H_0$. One finds
\begin{equation}
E_0=E_B+\frac{3}{5}N_F\epsilon_F \;,
\label{gse1}
\end{equation}
where $E_B$ is given by (\ref{bog4}) and
$\epsilon_F=\hbar^2 k_F^2/2m_F=\hbar^2(6\pi^2n_F)^{2/3}/2m_F$ is the Fermi energy, $k_F$ being the Fermi 
wave-vector. The mean-field term $\Delta E^{(1)}=g_{BF}N_B n_F$ provides
the first order correction to $E_0$. To second order in $g_{BF}$ one obtains
the result
\begin{eqnarray}
\Delta E^{(2)} &=& N_B g_{BF}^2 n_F\frac{1}{V}\sum_{\bf k} \frac{2m_R}{\hbar^2k^2} 
\nonumber\\
&+& \sum_{n\neq 0} 
\frac{|\langle 0|H_{int}|n\rangle|^2}{E_0-E_n} \;,
\label{gse2}
\end{eqnarray}
where $|n\rangle$ are the excited states of the unperturbed system with energy $E_n$.
The first term in the above equation arises from $\Delta E^{(1)}$ due to the
renormalization of the boson-fermion scattering length $g_{BF}\to g_{BF}+g_{BF}^2 1/V\sum_{\bf k} 2m_R/\hbar^2 k^2$.
The second term is the standard result of second order perturbation theory.
By directly calculating the matrix elements of the interaction Hamiltonian $H_{int}$, one finds the following result
for the ground-state energy
\begin{eqnarray}
E&=&E_0+N_Bg_{BF}n_F + N_Bg_{BF}^2n_F\frac{1}{V}\sum_{\bf k}\frac{2m_R}{\hbar^2 k^2} \nonumber \\
&-& N_Bg_{BF}^2\frac{1}{V^2}\sum_{{\bf k},{\bf q}}(u_k+v_k)^2
\frac{n_{\bf q}^F(1-n_{{\bf q}+{\bf k}}^F)}{\omega_k+\epsilon_{|{\bf q}+{\bf k}|}^F-\epsilon_q^F} \;,
\label{gse3}
\end{eqnarray}
where $n_{\bf q}^F=\theta(k_F-q)$ is the momentum distribution of the Fermi gas. We notice that the renormalization
of the boson-fermion scattering length is crucial in order to cancel the ultraviolet divergence in the integral 
over the wave-vector ${\bf k}$. The dependence on the relevant parameters becomes clearer by expressing the integrals 
over wave-vectors in units of $k_F$. Result (\ref{gse3}) takes the form
\begin{equation}
E=E_0+N_Bg_{BF}n_F + N_B\epsilon_F(n_Fb^3)^{2/3} A(w,\alpha) \;,
\label{gse4}
\end{equation}
where the second-order energy shift is expressed in terms of the Fermi energy $\epsilon_F$, the fermionic gas 
parameter $n_Fb^3$ and the dimensionless function
\begin{eqnarray}
&&A(w,\alpha) = \frac{2}{3}\left(\frac{6}{\pi}\right)^{2/3}\frac{1+w}{w}\int_0^\infty dk \int_{-1}^{+1} d\Omega
\Bigg[ 1
\nonumber\\
\label{gse5}\\
&-& \frac{3k^2(1+w)}{\sqrt{k^2+\alpha}}\int_0^1 dq q^2\frac{1-\theta(1-\sqrt{q^2+k^2+2kq\Omega})}
{\sqrt{k^2+\alpha}+wk+2qw\Omega}\Bigg].
\nonumber
\end{eqnarray}
In the above expression $w=m_B/m_F$ is the mass ratio and $\alpha=2/(k_F\xi_B)^2$ is a dimensionless parameter
fixed by the product of the Fermi wave-vector $k_F$ and the Bose healing length $\xi_B=1/\sqrt{8\pi n_0a}$.
In Fig. 1 we show the dependence of the function $A(w,\alpha)$ on the parameter $\alpha$ for a fixed value $w=1$ 
of the mass ratio. Two regimes are worth studying at this point. The regime $k_F\xi_B\gg1$ ($\alpha\ll 1$)
corresponds to a system where the Fermi energy is much larger than the chemical potential of the 
bosons $\epsilon_F\gg g_{BB}n_0$ (we assume $w\simeq 1$). In this regime, expression (\ref{gse5}) can be 
simplified as it depends only on the mass ratio $w$: $A(w,\alpha\to 0)=A_1(w)$, where 
\begin{eqnarray}
A_1(w)=\frac{2}{3}\left(\frac{6}{\pi}\right)^{2/3}\frac{1+w}{w}\int_0^\infty dk \int_{-1}^{+1} d\Omega
\Bigg[1
\nonumber\\ 
\label{gse6}\\
- 3k(1+w)\int_0^1 dq q^2\frac{1-\theta(1-\sqrt{q^2+k^2+2kq\Omega})}
{(1+w)k+2qw\Omega}\Bigg].
\nonumber
\end{eqnarray}
In this regime result (\ref{gse4}) reads
\begin{equation}
E=E_0+N_Bg_{BF}n_F + N_B\epsilon_F(n_Fb^3)^{2/3} A_1(w) \;,
\label{gse7}
\end{equation}
and coincides with the finding of Ref. \cite{AGIW} obtained using the T-matrix approach.
Notice that the effect of the Bose-Fermi interaction on the ground-state 
energy is independent of the Bose-Bose
interaction. In fact, in the limit $k_F\xi_B\gg 1$, the relevant contribution to the integral in Eq. (\ref{gse3})
comes from the large momentum region where the bosons behave as independent particles.
The opposite regime $k_F\xi_B\ll1$ ($\alpha\gg 1$) corresponds to a Fermi energy
$\epsilon_F\ll g_{BB}n_0$ ($w\simeq 1$). One finds: $A(w,\alpha\to \infty)=2\sqrt{\alpha}A_2(w)$, where
\begin{eqnarray}
A_2(w)&=& \frac{2}{3}\left(\frac{6}{\pi}\right)^{2/3}\frac{1+w}{w} \int_0^\infty dk 
\;\times 
\nonumber\\
&\times& \left[ 1 - \frac{(1+w) k^2}
{\sqrt{1+k^2}(\sqrt{1+k^2}+wk)}\right].
\label{gse8}
\end{eqnarray}
In this regime the ground-state energy of the system can be written as
\begin{eqnarray}
E&=&E_0+N_Bg_{BF}n_F 
\label{gse9}\\
&+& \frac{8}{6^{1/3}\pi^{1/6}}N_B\epsilon_F(n_Fb^3)^{2/3}
\left(\frac{n_B}{n_F}\right)^{1/3}(n_Ba^3)^{1/6} A_2(w) \;.
\nonumber
\end{eqnarray}
In contrast to result (\ref{gse7}), Bose-Bose interactions are important if $k_F\xi_B\ll1$ and the second order 
correction depends explicitly on the Bose gas parameter. Result (\ref{gse9}) has been first derived by Saam 
\cite{SAAM} for a dilute mixture with $a=b$. The dependence of $A_1$ and $A_2$ on the mass ratio $w$ is presented 
in Fig. 2.

\begin{figure}[htbp]
  \begin{center}
    \psfig{file=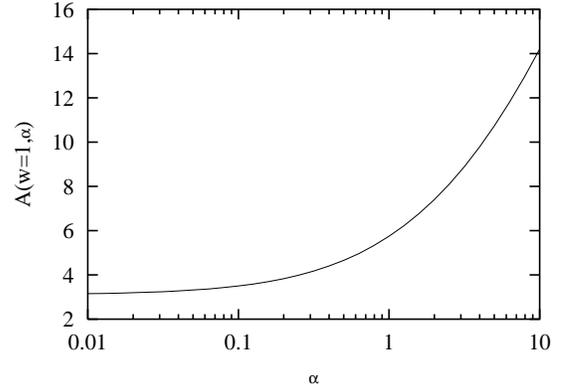,width=0.3\textwidth,angle=-90}
    \vspace{.2cm}
    \caption{Dimensionless function $A(w,\alpha)$ [Eq. (\ref{gse5})] as a function of $\alpha$ for the value
     $w=1$ of the mass ratio.}
    \label{fig1}
  \end{center}
\end{figure}

\begin{figure}[htbp]
  \begin{center}
    \psfig{file=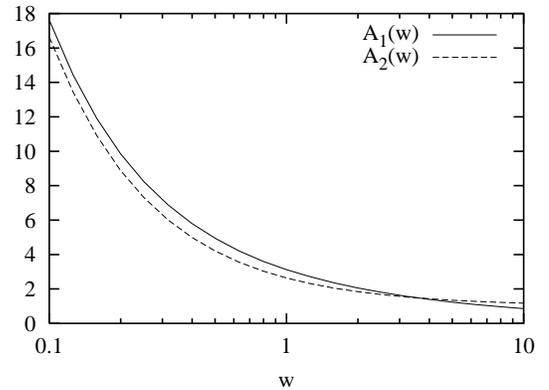,width=0.3\textwidth,angle=-90}
    \vspace{.2cm}
    \caption{Dimensionless functions $A_1(w)$ [Eq. (\ref{gse6})] (solid line) and $A_2(w)$ 
    [Eq. (\ref{gse8})] (dashed line).}
    \label{fig2}
  \end{center}
\end{figure}

\subsection{Phonon spectrum}

In this subsection we calculate the effect of fermions on the dispersion of the bosonic phonon branch. To this aim
it is convenient to use the Green's function formalism. The bosonic quasi-particle Green's function is defined as
\begin{equation}
D({\bf k},t)=-\frac{i}{\hbar}\langle T(\alpha_{\bf k}(t)\alpha_{\bf k}^\dagger(0))\rangle \;,
\label{gf1}
\end{equation}
where $T(...)$ is the time ordered product. Perturbation theory provides us with a precise recipe for calculating 
the Green's function $D({\bf k},t)$ to a given order in $H_{int}$
\begin{equation}
D({\bf k},t)=-\frac{i}{\hbar}\frac{1}{\langle S\rangle} 
\langle T(\alpha_{\bf k}(t)\alpha_{\bf k}^\dagger(0)S)\rangle \;,
\label{gf2}
\end{equation} 
where the time evolution operator $S$ is defined through the series expansion
\begin{eqnarray}
S=\sum_{n=0}^{\infty}\left(-\frac{i}{\hbar}\right)^n\frac{1}{n!}& &\int_{-\infty}^{+\infty}dt_1 ...
\int_{-\infty}^{+\infty}dt_n \; \times \nonumber \\
&\times& T\Big[H_{int}(t_1) ... H_{int}(t_n)\Big] \;.
\label{gf3}
\end{eqnarray}
In Fourier space the unperturbed quasi-particle Green's function is given by 
\begin{equation}
D_0({\bf k},\omega)=\frac{1}{\hbar\omega-\omega_k+i\eta} \;,
\label{gf4}
\end{equation} 
where $\eta>0$ and infinitesimally small. To second order in $H_{int}$ a direct calculation gives the result
\begin{eqnarray}
D({\bf k},\omega)&=&D_0({\bf k},\omega) \nonumber \\ 
&+& g_{BF}^2 n_0 (u_k+v_k)^2 D_0^2({\bf k},\omega) \Pi_0^F({\bf k},\omega)\;.
\label{gf5}
\end{eqnarray}
In the above equation $\Pi_0^F({\bf k},\omega)$ is the density-density response function of an ideal Fermi gas
defined as
\begin{eqnarray}
\Pi_0^F({\bf k},\omega)= \frac{1}{V}\sum_{\bf q} n_{\bf q}^F(1&-&n_{{\bf q}+{\bf k}}^F) \times   
\label{gf6} \\
\times \left( \frac{1}{\hbar\omega+\epsilon_q^F-\epsilon_{|{\bf q}+{\bf k}|}^F+i\eta} \right.&-&\left. 
\frac{1}{\hbar\omega+\epsilon_{|{\bf q}+{\bf k}|}^F-\epsilon_q^F-i\eta}\right) \;. \nonumber
\end{eqnarray} 
By solving Eq. (\ref{gf5}) for $D^{-1}({\bf k},\omega)$ one gets the following result for the excitation energy
to order $g_{BF}^2$
\begin{equation}
\hbar\omega=\omega_k+g_{BF}^2n_0\frac{\epsilon_k^B}{\omega_k}\Pi_0^F({\bf k},\hbar\omega=\omega_k) \;.
\label{gf7}
\end{equation}
The real and imaginary part of the fermionic response function $\Pi_0^F$ give rise respectively to a frequency 
shift and a damping of the quasi-particle. We are interested in the collective modes of the boson system which 
correspond to low momenta $\omega_k\ll g_{BB}n_0$ and $k\ll k_F$.
In the regime $k_F\xi_B\gg 1$ the fermionic density-density response function can be approximated by \cite{FET}
\begin{equation}
\Pi_0^F({\bf k},\hbar\omega=\omega_k)=-\frac{m_Fk_F}{2\pi^2\hbar^2}
\left(1+i\frac{\pi}{2\sqrt{2}w k_F\xi_B}\right) \;,
\label{gf8}
\end{equation}
and one gets the following result for the phonon excitation energy
\begin{eqnarray}
\hbar\omega&=&\hbar k c_B\Big[ 1-\frac{6^{1/3}(1+w)^2}{4\pi^{1/3}w}
\frac{b}{a} (n_F b^3)^{1/3} 
\nonumber\\
&-& i \frac{\sqrt{\pi}(1+w)^2}
{4w^2} \frac{b^2}{a^2} (n_Ba^3)^{1/2} \Big]\;.
\label{gf9}
\end{eqnarray}
In the above expression the real part in the square bracket gives the correction to the Bogoliubov sound velocity
$c_B=\sqrt{g_{BB}n_0/m_B}$, and the imaginary part gives the Landau damping of the phonon excitation due to 
collisions with the fermions. We notice that the correction to the sound velocity (second term in the square 
bracket) can be rewritten as $N(0)g_{BF}^2/2g_{BB}$, with $N(0)=m_Fk_F/(2\pi^2\hbar^2)$ being the density of 
states at the Fermi surface. The stability condition requiring that the real part of the excitation energy be
positive, corresponds to the linear stability condition $N(0)g_{BF}^2/g_{BB}<1$ (see Sec. III). In the regime 
$k_F\xi_B\gg 1$ the Bogoliubov phonon becomes soft when approaching the condition where the system phase 
separates. In the opposite regime $k_F\xi_B\ll 1$, the imaginary part of $\Pi_0^F$ vanishes 
at low momenta and the renormalized velocity of sound of the undamped phonon mode becomes
\begin{equation}
\hbar\omega=\hbar k c_B\Big[ 1 + \frac{w(1+w)^2}{8}\frac{b^2}{a^2}\frac{n_F}{n_B}\Big] \;.
\label{gf10}
\end{equation} 
The dispersion of phonons in dilute mixtures has also been 
discussed by Yip \cite{YIP} under more general conditions using
a numerical approach. He also predicts a shift 
towards lower frequencies with Landau damping for $k_F\xi_B\gg 1$, and
towards higher frequencies with no damping for $k_F\xi_B\ll 1$.

\subsection{Condensate fraction and momentum distribution}

The effect of the Bose-Fermi interaction on the momentum distribution of the bosons can be readily obtained 
by applying the perturbative approach to the bosonic single-particle Green's function 
\begin{equation}
G_{11}({\bf k},t)=-\frac{i}{\hbar}\langle T(a_{\bf k}(t)a_{\bf k}^\dagger(0))\rangle \;.
\label{cf1}
\end{equation}
The unperturbed single-particle Green's function is given by
\begin{eqnarray}
G_{11}^0({\bf k},\omega)&=&u_k^2D_0({\bf k},\omega)+v_k^2D_0({\bf k},-\omega) \nonumber \\
&=& \frac{u_k^2}{\hbar\omega-\omega_k+i\eta}-\frac{v_k^2}{\hbar\omega+\omega_k-i\eta} \;.
\label{cf2}
\end{eqnarray}
To second order in the Bose-Fermi coupling constant one gets the result
\begin{eqnarray}
G_{11}({\bf k},\omega)&=&G_{11}^0({\bf k},\omega) + g_{BF}^2n_0(u_k+v_k)^2 \Pi_0^F({\bf k},\omega)
\times \nonumber\\
&\times& \Big[ u_k^2D_0^2({\bf k},\omega) + v_k^2D_0^2({\bf k},-\omega) + 
\label{cf3}\\
&+&  2u_kv_k D_0({\bf k},\omega)D_0({\bf k},-\omega) \Big] \;.
\nonumber
\end{eqnarray}
The momentum distribution of the bosons can be obtained from the single-particle Green's function 
through the following relation
\begin{equation}
n_{\bf k}^B=\langle a_k^\dagger a_k\rangle=\frac{i\hbar}{2\pi}\int_{-\infty}^{+\infty}d\omega \;
G_{11}({\bf k},\omega) e^{i\omega\eta} \;,
\label{cf4}
\end{equation} 
which gives the result
\begin{eqnarray}
n_{\bf k}^B&=& v_k^2 + g_{BF}^2n_0(u_k+v_k)^2 \frac{1}{V}\sum_{\bf q} n_{\bf q}^F (1-n_{{\bf q}+{\bf k}}^F)
\times \nonumber\\
&\times& \left( \frac{u_k^2}{(\omega_k+\epsilon_{|{\bf k}+{\bf q}|}^F-\epsilon_q^F)^2} 
+  \frac{v_k^2}{(\omega_k+\epsilon_{|{\bf k}+{\bf q}|}^F-\epsilon_q^F)^2} \right.
\nonumber \\
&+&  \left. \frac{2u_k v_k}{\omega_k(\omega_k+\epsilon_{|{\bf k}+{\bf q}|}^F-\epsilon_q^F)} \right) \;.
\label{cf5}
\end{eqnarray}
By integrating $n_{\bf k}^B$ over momenta one obtains the fraction of atoms which are scattered out of the 
condensate due to interaction. The quantum depletion of the condensate is given by
\begin{eqnarray}
\frac{N^\prime}{N_B}&=&\frac{8}{3\sqrt{\pi}}(n_Ba^3)^{1/2} + g_{BF}^2\frac{1}{V}\sum_{\bf k}
\frac{\epsilon_k^B}{\omega_k} \frac{1}{V}\sum_{\bf q} \times
\label{cf6}\\
&\times& \frac{n_{\bf q}^F (1-n_{{\bf q}+{\bf k}}^F)}{(\omega_k+\epsilon_{|{\bf k}+{\bf q}|}^F-\epsilon_q^F)^2}
\left(\frac{\epsilon_k^B}{\omega_k}-\frac{g_{BB}n_0}{\omega_k^2}(\epsilon_{|{\bf k}+{\bf q}|}^F-\epsilon_q^F)
\right) .
\nonumber
\end{eqnarray}
The first term in the above equation corresponds to the Bogoliubov depletion present also in a pure bosonic 
system which
is due to interaction effects among the bosons. The second term accounts instead for boson-fermion scattering 
processes. The condensate fraction is obtained from the difference $N_0/N_B=1-N^\prime/N_B$. By writing the 
integrals over wave-vectors in units of $k_F$, result (\ref{cf6}) reads
\begin{equation}
\frac{N^\prime}{N_B}=\frac{8}{3\sqrt{\pi}}(n_Ba^3)^{1/2} + (n_Fb^3)^{2/3} B(w,\alpha) \;.
\label{cf7}
\end{equation}
As for result (\ref{gse4}), the dimensionless function $B$ depends on the mass ratio $w=m_B/m_F$ and on the 
parameter $\alpha=2/(k_F\xi_B)^2$ and is given by
\begin{eqnarray}
&&B(w,\alpha) = \frac{12}{6^{1/3}\pi^{2/3}} (1+w)^2 \int_0^\infty dk \int_{-1}^{+1} d\Omega \times
\nonumber\\
&\times& \frac{k^2}{\sqrt{k^2+\alpha}} \int_0^1 dq q^2\frac{1-\theta(1-\sqrt{q^2+k^2+2kq\Omega})}
{(\sqrt{k^2+\alpha}+wk+2qw\Omega)^2} \times
\nonumber\\
&\times& \left(\frac{1}{\sqrt{k^2+\alpha}}-\frac{\alpha(wk+2qw\Omega)}{2k^2(k^2+\alpha)}\right) \;.
\label{cf8}
\end{eqnarray} 
The dependence of the function $B$ on the parameter $\alpha$ is shown in Fig. 3 for the mass ratio $w=1$.
In the regime $k_F\xi_B\gg 1$ ($\alpha\ll 1$) the quantum depletion due to scattering processes with the fermions 
becomes independent of the Bose-Bose coupling constant and one finds
\begin{equation}
\frac{N^\prime}{N_B}=\frac{8}{3\sqrt{\pi}}(n_Ba^3)^{1/2} + (n_Fb^3)^{2/3} B_1(w) \;,
\label{cf9}
\end{equation}
where the function $B_1$ of the mass ratio is given by
\begin{eqnarray}
B_1(w) &=& \frac{12}{6^{1/3}\pi^{2/3}}(w+1) \int_{-1}^{+1} d\Omega 
\int_0^1 dq q^2 \times\\
\nonumber
&\times&\frac{1}{(w+1)\sqrt{q^2(\Omega^2-1)+1}
+(w-1)q\Omega}\;.
\label{cf10}
\end{eqnarray} 
In the opposite regime $k_F\xi_B\ll 1$ ($\alpha\gg 1$), the quantum depletion takes important contributions
from the region of low momenta $k<mc_B/\hbar$ and interaction effects among the bosons become crucial.
In this regime one finds: $B(w,\alpha\to\infty)=2B_2(w)/(3\sqrt{\alpha})$ where the function $B_2$ is given by
\begin{eqnarray}
B_2(w) &=& \frac{12}{6^{1/3}\pi^{2/3}}(1+w)^2 \int_0^\infty dk \frac{k^2}{1+k^2} \times
\nonumber\\
&\times& \frac{1}{(\sqrt{1+k^2}+wk)^2}\left(1-\frac{w}{2k\sqrt{1+k^2}}\right) \,
\label{cf11}
\end{eqnarray} 
and the quantum depletion takes consequently the form
\begin{eqnarray}
\frac{N^\prime}{N_B}&=&\frac{8}{3\sqrt{\pi}}(n_Ba^3)^{1/2} 
\nonumber\\
&+& \frac{\pi^{1/6}}{6^{2/3}} \frac{b^2}{a^2}
(n_Ba^3)^{1/2}\frac{n_F}{n_B} B_2(w) \;.
\label{cf12}
\end{eqnarray}
In Fig. 4 we plot the functions $B_1$ and $B_2$ of the mass ratio $w$. We notice that, while $B_1$ is 
always positive, $B_2$ is positive for small values of $w$ and becomes large and negative when $m_B\gg m_F$.
Thus, if the fermions are light enough compared to the bosons, their effect in the regime $k_F\xi_B\ll 1$ is 
to enhance the occupation of the condensate state resulting in a decrease of the quantum depletion. 
Of course, when $B_2$ is negative, result (\ref{cf12}) is valid only if the second term is small compared
to the first one. 

\begin{figure}[htbp]
  \begin{center}
    \psfig{file=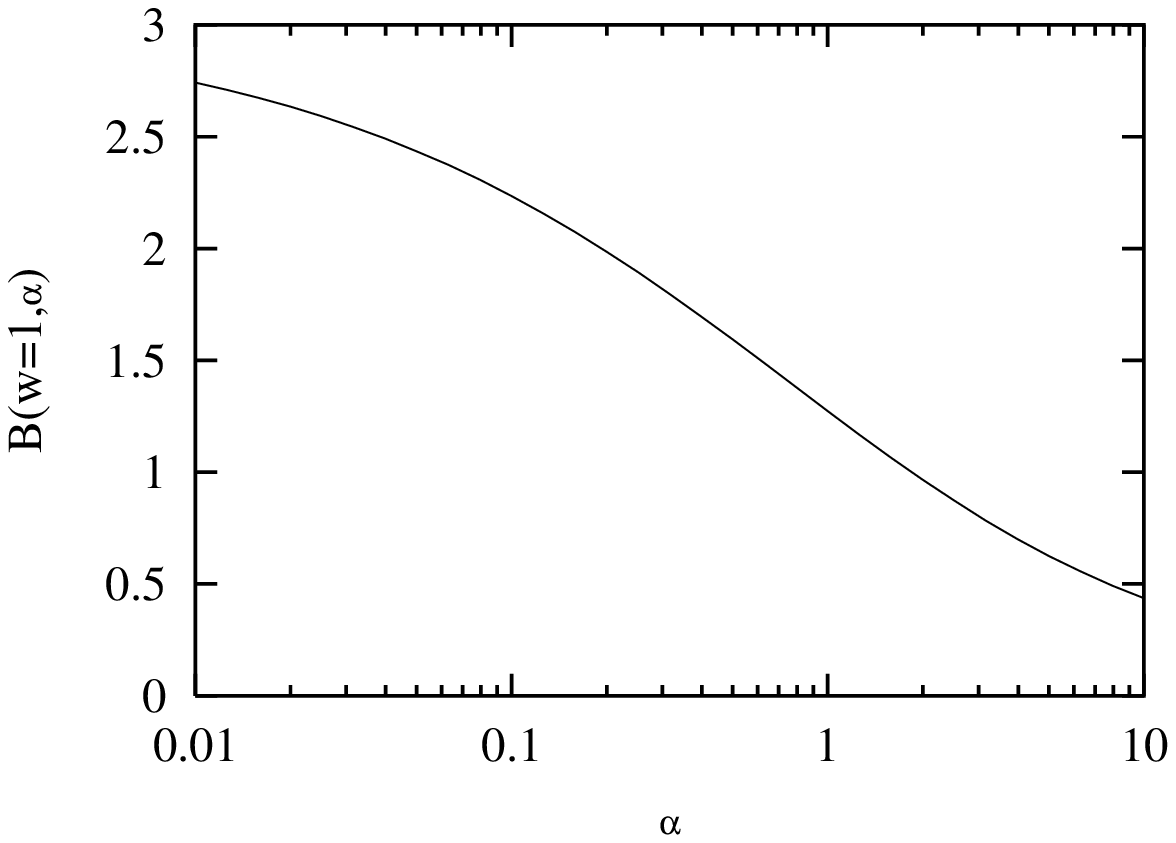,width=0.3\textwidth,angle=-90}
    \vspace{.2cm}
    \caption{Dimensionless function $B(w,\alpha)$ [Eq. (\ref{cf8})] as a function of $\alpha$ for the value
     $w=1$ of the mass ratio.}
    \label{fig3}
  \end{center}
\end{figure}

\begin{figure}[htbp]
  \begin{center}
    \psfig{file=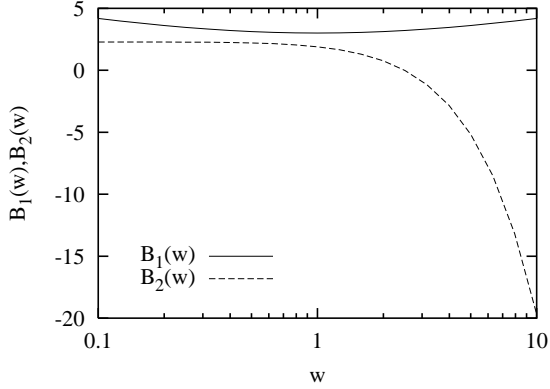,width=0.3\textwidth,angle=-90}
    \vspace{.2cm}
    \caption{Dimensionless functions $B_1(w)$ [Eq. (\ref{cf10})] (solid line) and $B_2(w)$ 
    [Eq. (\ref{cf11})] (dashed line).}
    \label{fig4}
  \end{center}
\end{figure}

Another quantity of interest is the momentum distribution of the bosons given by eq. (\ref{cf5}).  For a 
pure system the momentum distribution is given by the well-known Bogoliubov result
\begin{equation}
n_{\bf k}^B|_{pure} = \frac{x^2+1-x\sqrt{x^2+2}}{2x\sqrt{x^2+2}} \;,
\label{cf13}
\end{equation}
which has been written in units of the inverse healing length $x=k\xi_B$. At small momenta $x\ll 1$, the momentum
distribution is dominated by the quantum fluctuations of phonons which give rise to the infrared divergence 
$n_{\bf k}^B|_{pure}\sim 1/\sqrt{8}x$. At large momenta
$x\gg 1$, one finds instead the algebraic decay 
$n_{\bf k}^B|_{pure}\sim 1/4x^4$. The important question arises concerning the effect of the fermionic component on 
$n_{\bf k}^B$. In units of the inverse healing length  result (\ref{cf5}) reads
\begin{eqnarray}
n_{\bf k}^B &=& n_{\bf k}^B|_{pure} + \frac{4(1+w)^2}{(6\pi^2)^{1/3}} (n_Fb^3)^{2/3} 
\frac{n_B}{n_F} \times 
\nonumber\\
&\times& \frac{k_F\xi_B}{\sqrt{x^2+2}}\int_{-1}^{+1} d\Omega 
\int_0^{k_F\xi_B} dy y^2 \times
\nonumber\\
&\times& \frac{1-\theta(k_F\xi_B-\sqrt{y^2+x^2+2xy\Omega})}
{(\sqrt{x^2+2}+wx+2yw\Omega)^2} \times
\label{cf14}\\
&\times& \left(\frac{1}{\sqrt{x^2+2}}-\frac{wx+2yw\Omega}{x^2(x^2+2)}\right) \;.
\nonumber
\end{eqnarray}
In a mixture there are two characteristic wavenumbers 
in the problem: $\xi_B^{-1}$ and $k_F$. 
The structure of the distribution is consequently
richer and depends on how $\xi_B^{-1}$ compares with $k_F$. Nevertheless
very interesting conclusions can be drawn in the two limits
$k\ll\xi_B^{-1},k_F$ and $k\gg\xi_B^{-1},k_F$.
At low momenta ($k\ll\xi_B^{-1},k_F$) the effect of the 
Bose-Fermi coupling is to enhance the occupation of the condensate 
by stimulating scattering of particles from ${\bf k}\neq 0$ states 
into the condensate. The contribution to the 
momentum distribution is thus negative and reduces the coefficient 
of the $1/x$ divergent term
\begin{eqnarray}
n_{\bf k}^B &\sim& \frac{1}{\sqrt{8}x} 
\Big[1-\frac{6^{1/3}(1+w)^2}{4\pi^{1/3}w}\frac{b}{a}
(n_Fb^3)^{1/3}\times
\nonumber\\
&\times& f(2w/\sqrt{\alpha})\Big]\;,
\label{cf15}
\end{eqnarray}
where
\begin{equation}
f(t)=\frac{2+t}{1+t}-\frac{2}{t}\ln(1+t)\;.
\end{equation}
Notice that the modulus of the Bose-Fermi contribution can be written as 
$N(0)g_{BF}^2/2g_{BB}\cdot f(2w/\sqrt{\alpha})$ [see comment after Eq. (\ref{gf9})].
Phase stability of the mixture requires that in any case
$N(0)g_{BF}^2/g_{BB}<1$. Since moreover $0<f(t)<1$ for all $t>0$,
we conclude that the Bose-Fermi term is always less, in modulus, than the
Bose-Bose one and thus Eq. (\ref{cf15}) is well defined.
It is also worth noticing that in the regime $k_F\xi_B\gg 1$ the function $f\to 1$
and $n_{\bf k}^B \sim (c/c_B)/\sqrt{8}x$, where $c$ is the renormalized sound velocity 
from Eq. (\ref{gf9}).

At large momenta ($k\gg\xi_B^{-1},k_F$)
the Bose-Fermi coupling 
provides instead further depletion of the condensate and one finds 
\begin{equation}
n_{\bf k}^B \sim \frac{1}{4x^4} 
\Big[1+\frac{32}{3(6\pi^2)^{1/3}}(n_Fb^3)^{2/3}\frac{n_B}{n_F}
(k_F\xi_B)^4 \Big] \;.
\label{cf16}
\end{equation}
In the regime $k_F\xi_B\gg 1$ the shape of $n_{\bf k}^B$ can be very 
different compared to the pure case. In fact,
the cut-off momentum increases from $\xi_B^{-1}$ to $k_F$ and the 
momentum distribution acquires a long tail at
large momenta (see Fig. 5). The effect of the Bose-Fermi interaction on the momentum
distribution and condensate fraction of the bosons has been calculated using variational 
methods in the hypernetted-chain scheme for the strongly correlated liquid $^3$He - $^4$He 
mixture \cite{HE34}. However, weak effects are found due to the low $^3$He concentration 
attainable in stable mixtures.

\subsection{Superfluid density}

An important aspect of Bose-Fermi mixtures concerns 
their superfluid behavior. In a pure bosonic system at zero temperature 
the density of the normal component vanishes, $\rho_n=0$, and 
the system is completely superfluid. In a Bose-Fermi
mixture the normal density does not vanish and can be written as $\rho_n=m_F^\ast n_F$ in terms of the effective 
mass
$m_F^\ast$ of the fermionic particles. In the absence of Bose-Fermi coupling, the effective mass coincides with 
the bare mass of the fermions, $m_F^\ast=m_F$, and the normal density of the mixture coincides with the mass density
of the fermionic component. Interaction effects dress the fermions resulting in an effective mass larger than the 
bare
mass. A well known example is provided by liquid $^3$He-$^4$He solutions, where due to the strong correlations the 
effective mass of $^3$He atoms 
is $m_F^\ast/m_F\simeq 2.3$ \cite{EDW}. An equivalent way of interpreting the excess of normal density in Bose-Fermi
mixtures is to write $\rho_n=m_F n_F+\rho_n^B$, where $\rho_n^B$ is the contribution arising from the excitation of
bosonic quasi-particles due to the Bose-Fermi coupling. In terms of the mass difference $\delta m_F=m_F^\ast-m_F$,
the bosonic component of $\rho_n$ is given by $\rho_n^B=n_F \delta m_F$.     
The superfluid density of the system is defined as $\rho_s=\rho_B-\rho_n^B$.
The bosonic component of the normal density can be calculated through the long-wavelength behavior of the static
transverse current-current response function \cite{BAYM}
\begin{equation}
\rho_n^B=-\lim_{{\bf q}\to 0} G_T({\bf q},\omega=0) \;.
\label{sd1}
\end{equation}
In the above equation $G_T({\bf q},t)=-i/\hbar\langle T(j_{\bf q}^x(t)j_{{-\bf q}}^x(0))\rangle$ is the 
bosonic transverse 
current-current response function, with ${\bf q}=(0,0,q)$ directed along $z$ and the $x$ component of
the current operator defined by $j_{\bf q}^x=1/\sqrt{V}\sum_{\bf k} \hbar k_x a_{\bf k}^\dagger 
a_{{\bf k}+{\bf q}}$. In terms of single-particle Green's functions $G_T({\bf q},\omega)$ can be written as
\begin{eqnarray}
G_T({\bf q},\omega) &=& \frac{i\hbar}{2\pi}\frac{1}{V}
\sum_{\bf k}(\hbar k_x)^2 \int_{-\infty}^{+\infty} 
d \omega^\prime \; \times 
\nonumber\\
&\times& \Big[ G_{11}({\bf k},-\omega^\prime)G_{11}({\bf k}+{\bf q},\omega-\omega^\prime) 
\nonumber\\ 
&-&  G_{21}({\bf k},\omega^\prime)G_{12}({\bf k}+{\bf q},\omega-\omega^\prime) \Big] \;,
\label{sd2}
\end{eqnarray}
where $G_{12}({\bf k},t)=-i/\hbar\langle T(a_{\bf k}(t)a_{{-\bf k}}(0))\rangle$ and 
$G_{21}({\bf k},t)=-i/\hbar\langle T(a_{\bf k}^\dagger(t)a_{{-\bf k}}^\dagger(0))\rangle$ are the 
anomalous single-particle Green's functions, which are peculiar of Bose condensed systems.
The unperturbed anomalous single-particle Green's functions can be readily obtained and one finds
\begin{eqnarray}
G_{12}^0({\bf k},\omega)&=&G_{21}^0({\bf k},\omega)
\nonumber\\
&=&\frac{u_k v_k}{\hbar\omega-\omega_k+i\eta}
- \frac{u_k v_k}{\hbar\omega+\omega_k-i\eta} \;.
\label{sd3}
\end{eqnarray}
To second order in $g_{BF}$ we find the result 
\begin{eqnarray}
G_{12}({\bf k},\omega)&=&G_{21}({\bf k},\omega)
\nonumber\\
&=&G_{12}^0({\bf k},\omega) + g_{BF}^2n_0(u_k+v_k)^2 \Pi_0^F({\bf k},\omega)
\times \nonumber\\
&\times& \Big[ (u_k^2+v_k^2) D_0({\bf k},\omega)D_0({\bf k},-\omega) +  
\label{sd4}\\
&+&  u_kv_k D_0^2({\bf k},\omega) + u_kv_k D_0^2({\bf k},-\omega) \Big] \;.
\nonumber
\end{eqnarray}
>From the perturbation expansions (\ref{cf3}) and (\ref{sd4}) one gets the following result for the 
long-wavelength limit of the static current-current response function
\begin{eqnarray}
\lim_{{\bf q}\to 0} G_T({\bf q},\omega=0) &=& 
\frac{i\hbar}{2\pi}\frac{2}{V}\sum_{\bf k}(\hbar k_x)^2 g_{BF}^2 n_0 (u_k+v_k)^2  \times
\nonumber\\
&\times&  \int_{-\infty}^{+\infty} 
d \omega\;  \Pi_0^F({\bf k},\omega) D_0^3({\bf k},\omega) \;,
\label{sd5}
\end{eqnarray}
valid to order $g_{BF}^2$. The fraction of Bose particles which contribute to the normal density of the 
system is thus given by the following expression
\begin{eqnarray}
\frac{\rho_n^B}{\rho_B}&=&\frac{4}{3} g_{BF}^2\frac{1}{V}\sum_{\bf k}
\frac{(\epsilon_k^B)^2}{\omega_k} \times
\nonumber\\
&\times& \frac{1}{V}\sum_{\bf q} 
\frac{n_{\bf q}^F (1-n_{{\bf q}+{\bf k}}^F)}{(\omega_k+\epsilon_{|{\bf k}+{\bf q}|}^F-\epsilon_q^F)^3}
\; .
\label{sd6}
\end{eqnarray}
By expressing the integrals in units of the Fermi wave-vector $k_F$ one finds the result
\begin{equation}
\frac{\rho_n^B}{\rho_B}= (n_Fb^3)^{2/3} C(w,\alpha) \;,
\label{sd7}
\end{equation}
in terms of the Fermi gas parameter and of the dimensionless function 
\begin{eqnarray}
&&C(w,\alpha) = \frac{16}{6^{1/3}\pi^{2/3}} (1+w)^2 \int_0^\infty dk \int_{-1}^{+1} d\Omega \times
\nonumber\\
&\times& \frac{k^2}{\sqrt{k^2+\alpha}} \int_0^1 dq q^2\frac{1-\theta(1-\sqrt{q^2+k^2+2kq\Omega})}
{(\sqrt{k^2+\alpha}+wk+2qw\Omega)^3} \;.
\label{sd8}
\end{eqnarray}
In Fig. 5 we show the dependence of the above function on the parameter $\alpha=2/(k_F\xi_B)^2$ for the 
value $w=1$ of the mass ratio.
In the regime $k_F\xi_B\gg 1$ the function $C$ becomes independent of the parameter $\alpha$ and result 
(\ref{sd7}) takes the form
\begin{equation}
\frac{\rho_n^B}{\rho_B}= (n_Fb^3)^{2/3} C_1(w) \;,
\label{sd9}
\end{equation} 
where $C_1$ is given by
\begin{eqnarray}
C_1(w) &=& \frac{16}{6^{1/3}\pi^{2/3}}\int_{-1}^{+1}
d\Omega \int_0^1 dq q^2\times\\
\nonumber
&\times& \frac{(1+w)\sqrt{q^2(\Omega^2-1)+1}-q\Omega}
{[(1+w)\sqrt{q^2(\Omega^2-1)+1}+(w-1)q\Omega]^2} \;.
\label{sd10}
\end{eqnarray}
In the opposite regime, $k_F\xi_B\ll 1$, the function $C$ takes the following asymptotic value 
$C(w,\alpha\to\infty)=2C_2(w)/(3\sqrt{\alpha})$ and one finds
\begin{equation}
\frac{\rho_n^B}{\rho_B}= \frac{\pi^{1/6}}{6^{2/3}} \frac{b^2}{a^2}
(n_Ba^3)^{1/2}\frac{n_F}{n_B} C_2(w) \;, 
\label{sd11}
\end{equation}   
where the function $C_2$ is given by
\begin{equation}
C_2(w) = \frac{16(1+w)^2}{6^{1/3}\pi^{2/3}}\int_0^\infty dk \frac{k^2}{\sqrt{1+k^2}(\sqrt{1+k^2}+wk)^3} 
\;.
\label{sd12}
\end{equation}
Result (\ref{sd11}-\ref{sd12}) is in agreement with the finding of Ref. \cite{SAAM} where the fermion 
effective mass has been calculated for $k_F\xi_B\ll 1$ in the case of equal scattering 
lengths $a=b$. In Fig. 6 we show both $C_1$ and $C_2$ as a function of the mass ratio.  

\begin{figure}[htbp]
  \begin{center}
    \psfig{file=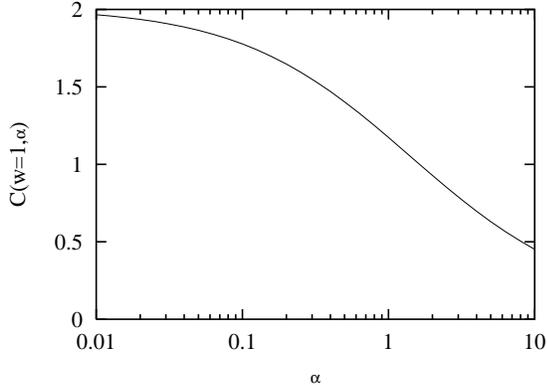,width=0.3\textwidth,angle=-90}
    \vspace{.2cm}
    \caption{Dimensionless function $C(w,\alpha)$ [Eq. (\ref{sd8})] as a function of $\alpha$ for the value
     $w=1$ of the mass ratio.}
    \label{fig5}
  \end{center}
\end{figure}

\begin{figure}[htbp]
  \begin{center}
    \psfig{file=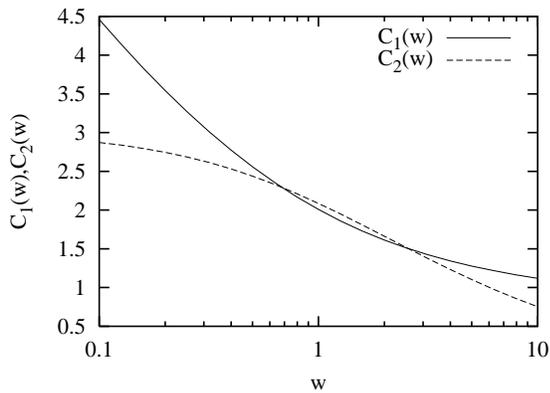,width=0.3\textwidth,angle=-90}
    \vspace{.2cm}
    \caption{Dimensionless functions $C_1(w)$ [Eq. (\ref{sd10})] (solid line) and $C_2(w)$ 
    [Eq. (\ref{sd12})] (dashed line).}
    \label{fig6}
  \end{center}
\end{figure}

\section{NUMERICAL RESULTS}

In deriving the above results we have been completely general.
In this section instead, we look at their implications on actual physical
systems and we discuss more quantitatively the striking consequences of the 
presence of fermions on the Bose-Einstein condensate which we 
have pointed out. 

First of all we have to
spend a few words on phase stability. The obvious requirement
for all of the above to be valid is that bosons and fermions mix
in the first place. According to the findings of ref. \cite{VIVPS}
one has to separately analyze two cases: attractive Bose-Fermi 
interaction, i.e. $g_{BF}<0$, and repulsive one, i.e. $g_{BF}>0$.
The former case is simpler, as the only requirement is that arising 
from linear stability: $N(0)g_{BF}^2/g_{BB}<1$, where 
$N(0)=m_Fk_F/(2\pi^2\hbar^2)$ is the density of states at the Fermi surface.  
This condition fixes an upper limit on the fermionic density irrespective 
of the bosonic one 
\begin{equation}
\label{nfmax}
n_{F,max}=\frac{4\pi}{3}\frac{w^3}{(1+w)^6}\left(\frac{a}{b}\right)^3
\frac{1}{b^3} \;.
\end{equation}
If instead $g_{BF}$ is greater than zero, while the fermionic
density cannot in any case be larger than $n_{F,max}$, yet for every
$n_F<n_{F,max}$ there is a critical $n_{B,c}(n_F)$ at which the system 
phase separates anyway. The function $n_{B,c}(n_F)$ is non trivial, and one
has to check specifically for a given pair of elements and densities
whether the system is stable or not.

For this reason it is certainly easier to observe important effects in a
mixture where $g_{BF}$ is negative, as one can reach higher densities
without running into problems. On the other hand, since our
calculations are to order $g_{BF}^2$ all effects computed here are identical 
for both signs of $g_{BF}$.

A mixture with negative $g_{BF}$, and 
supposingly large $b$, is already available experimentally.
This is the $^{40}$K-$^{87}$Rb mixture recently reported in 
ref. \cite{ROATI}. According to the authors $b=300(100)a_0$
in units of the Bohr radius. 

We consider a mixture of $^{87}$Rb and $^{40}$K with densities $n_B=n_F=3\cdot 10^{14}$ 
cm$^{-3}$. For the scattering lengths we use $a=$110$a_0$ and $b=$300$a_0$, 
and the mass ratio is $w=$2.175. The value of the parameter $k_F\xi_B$ is 3.9. 

In Fig. 7 we show results for the boson momentum distribution of the mixture, 
obtained from Eq. (\ref{cf14}) with the parameters given above, compared with
the one of a pure system [eq. (\ref{cf13})]. The total value of the quantum 
depletion is $\sim$ 3\% for this configuration and the Bose-Bose and Bose-Fermi 
contributions are $\sim$ 1\% and $\sim$ 2\% respectively (see Fig. 8). 
We see a large effect due to
boson-fermion interactions. The momentum distribution at low momenta is 
depressed while at large momenta is considerably enhanced and a tail appears
which extends to very large momenta. 

In Fig. 8 we report results on the dependence of the condensate depletion and 
normal fluid fraction as a function of the fermion density $n_F$. The boson 
density has been fixed to $3\cdot 10^{14}$ cm$^{-3}$. With the above values of the 
parameters the system becomes unstable at a fermion density of 
$5.2\cdot10^{14}$ cm$^{-3}$. We see that, by increasing the fermion density,
both the condensate depletion and the normal fluid fraction increase. The
prediction for fermion densities close to the instability threshold gives 
$N^\prime/N_B\sim$ 4\% and $\rho_n^B/\rho_B\sim$ 2\%.  

In Fig. 9 we show the condensate depletion and the normal fluid fraction of a 
hypothetical mixture with mass ratio $m_B/m_F=0.1$. We also assume $g_{BF}<0$
and we have fixed the scattering lengths to $a=100a_0$ and $b=50a_0$, and the 
boson density to $n_B=3\cdot 10^{14}$ cm$^{-3}$. Such system becomes unstable for 
$n_F\geq 1.0\cdot 10^{18}$ cm$^{-3}$. We see that an increase of $n_F$ is followed 
by an increase of both the condensate depletion and the normal fluid fraction.
However, at the fermion density $n_F\sim 2\cdot 10^{17}$ cm$^{-3}$, 
the normal fluid fraction becomes larger
than the condensate depletion. This effect has been first discussed in connection with
disordered dilute Bose gases by Huang and Meng \cite{HM} and very recently has been 
explicitly proved in a Bose gas with quenched impurities using quantum Monte-Carlo 
techniques \cite{AG}. It is striking that the same effect is found for Bose-Fermi 
mixtures in regimes of high fermion concentration when $m_F\gg m_B$.

Finally, in Fig. 10 we present results of the condensate depletion and normal fluid fraction
of a different hypothetical mixture with mass ratio $m_B/m_F=20$. We assume again an
attractive Bose-Fermi coupling ($g_{BF}<0$) and we use the following parameters:
$a=100a_0$, $b=150a_0$ and $n_B=3\cdot 10^{14}$ cm$^{-3}$. The mixture becomes unstable for 
$n_F\geq 2.3\cdot 10^{14}$ cm$^{-3}$. In this case we see a different striking behavior:
for low fermion concentration the effect of the Bose-Fermi coupling is to stimulate the 
occupation of the condensate, resulting in a quantum depletion smaller than in the pure
case. 

Due to the smallness of the effects, the interest of the results reported in 
Figs. \ref{fig9} and \ref{fig10} is to show
that they can actually occur in realistic mixtures, though
it might be difficult to observe them with present techniques.

\section{CONCLUSIONS}
We have studied the properties of a dilute Bose-Fermi mixture at zero temperature using a 
perturbation approach. We have investigated both equilibrium properties, such as the ground-state
energy, the boson momentum distribution and the normal fluid fraction, and dynamic properties such 
as the dispersion law and damping of phonon excitations. The system is very rich. By varying the mass
ratio, the densities and the scattering lengths of the two components one can obtain very different 
regimes where striking effects due to boson-fermion interactions can occurr. These include: the strong 
suppression of the boson momentum distribution at low momenta, localization effects exhibited by the superfluid 
density
which becomes smaller than the condensate fraction and stimulated scattering of bosons into the 
condensate.

\section{ACKNOWLEDGMENTS}
We gratefully acknowledge useful discussions with L.P. Pitaevskii and S. Stringari. This research is 
supported by Ministero dell'Istruzione, dell'Universit\`a e della Ricerca (MIUR).

\begin{figure}[htbp]
  \begin{center}
    \psfig{file=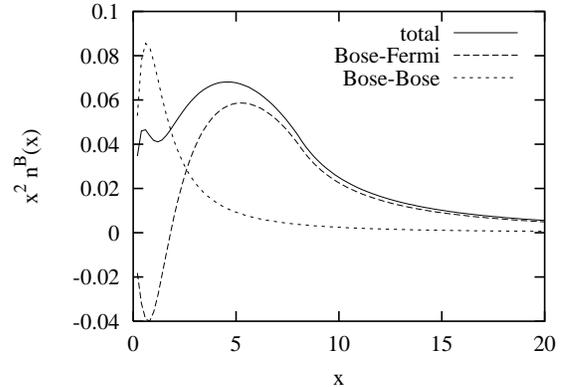,width=0.3\textwidth,angle=-90}
    \vspace{.2cm}
    \caption{Boson momentum distribution of a $^{87}$Rb -$^{40}$K mixture 
     (solid line). The momentum distribution of a pure system (short dashed) 
      and the Bose-Fermi contribution (long dashed) are also shown. 
      In the figure as in the text $x$ stands for $k\xi_B$.}
    \label{fig7}
  \end{center}
\end{figure}

\begin{figure}[htbp]
  \begin{center}
    \psfig{file=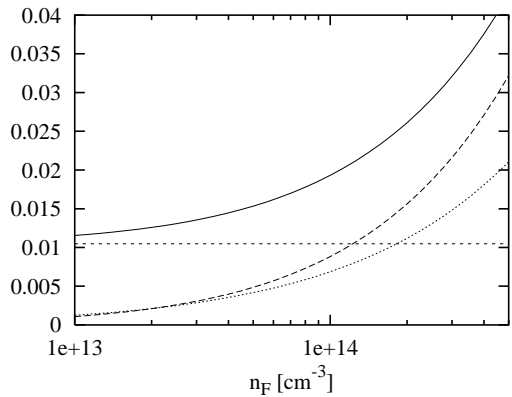,width=0.3\textwidth,angle=-90}
    \vspace{.2cm}
    \caption{Condensate depletion and normal fluid fraction of a $^{87}$Rb-$^{40}$K 
    mixture as a function of the fermion density. The total condensate depletion is shown with 
    the solid line. The Bogoliubov and the fermion depletion are shown respectively with the 
    horizontal short-dashed line and the long-dashed line. The dotted line corresponds instead 
    to the normal fluid fraction.} 
    \label{fig8}
  \end{center}
\end{figure}

\begin{figure}[htbp]
  \begin{center}
    \psfig{file=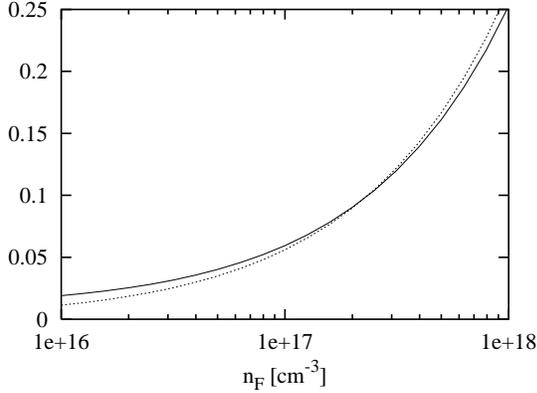,width=0.3\textwidth,angle=-90}
    \vspace{.2cm}
    \caption{Condensate depletion (solid line) and normal fluid fraction (dotted line) of a 
    hypothetical mixture with mass ratio $m_B/m_F$=0.1 as a function of the fermion density.}
    \label{fig9}
  \end{center}
\end{figure}

\begin{figure}[htbp]
  \begin{center}
    \psfig{file=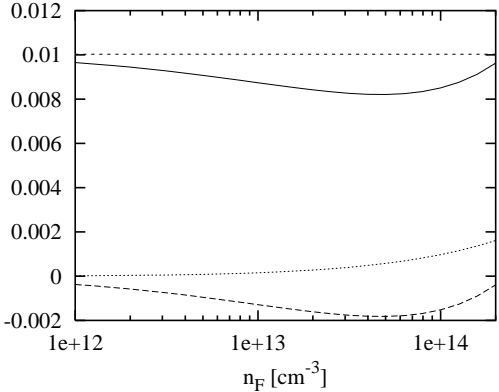,width=0.3\textwidth,angle=-90}
    \vspace{.2cm}
    \caption{Condensate depletion and normal fluid fraction of a 
    hypothetical mixture with mass ratio $m_B/m_F=$20 as a function of the fermion density.
    The line codes used are the same as for Fig. 8.}
    \label{fig10}
  \end{center}
\end{figure}

\end{document}